\documentclass[journal]{IEEEtran}

\usepackage{amsmath,epsfig,bm,amssymb,amsthm}
\usepackage{psfrag}
\usepackage{cite}

\ifCLASSINFOpdf
\else
\fi

\usepackage{amsmath,epsfig,bm,amssymb,amsthm,balance}

\newcommand{\dd}{{\mathrm d}}
\newcommand{\sd}{\mathrm{sd}}
\newcommand{\rd}{\mathrm{rd}}
\newcommand{\sr}{\mathrm{sr}}

\newcommand{\gth}{\gamma_{\mathrm{th}}}

\begin{document}
%
\title{Selection Combining for Differential Amplify-and-Forward Relaying Over Rayleigh-Fading Channels
\thanks{The authors are with the Department of Electrical and Computer Engineering,
University of Saskatchewan, Saskatoon, Canada, S7N5A9.
Email: m.avendi@usask.ca, ha.nguyen@usask.ca.}}

\author{\IEEEauthorblockN{M. R. Avendi and Ha H. Nguyen}
}

\maketitle

\begin{abstract}
\label{abs}
This paper proposes and analyses selection combining (SC) at the destination for differential amplify-and-forward (D-AF) relaying over slow Rayleigh-fading channels. The selection combiner chooses the link with the maximum magnitude of the decision variable to be used for non-coherent detection of the transmitted symbols. Therefore, in contrast to the maximum ratio combining (MRC), no channel information is needed at the destination. The exact average bit-error-rate (BER) of the proposed SC is derived and verified with simulation results. It is also shown that the performance of the SC method is very close to that of the MRC method, albeit with lower complexity.
\end{abstract}


\IEEEpeerreviewmaketitle

\section{Introduction}
\label{se:intro}
The idea of employing other wireless users as relays in a communication network was proposed more than a decade ago \cite{uplink-Aazgang}. Cooperative communication exploits the fact that, since other users can also listen to a source, they would be able to receive, process and re-broadcast the received data to the destination. Depending on the strategy that relays utilize for cooperation, the relay networks are generally classified as decode-and-forward (DF) and amplify-and-forward (AF)\cite{coop-laneman}.

Among these two strategies, AF is very attractive in terms of having less computational burden at the relays. Specifically, the relay's function is simply to multiply the received signal with a fixed or variable amplification factor, depending on the availability of the channels state information (CSI). In the case of having no CSI at the relays, the second-order statistics of source-relay channels can be used to determine a fixed amplification factor. Also, using differential encoding, differential AF (D-AF) scheme has been considered in \cite{DAF-Liu,DAF-DDF-QZ,DAF-General} to avoid channel estimation at the destination. In the absence of CSI at the destination, a set of fixed weights, based on the second-order statistics of all channels, have been used to combine the received signals from the relay-destination and the source-destination links \cite{DAF-Liu,DAF-DDF-QZ,DAF-General}. For future reference, this combiner is called semi-maximum ratio combining (semi-MRC). Since the exact performance analysis of semi-MRC appears to be too complicated (if not impossible), the performance of a system using instantaneous combining weights (i.e., the instantaneous MRC) is usually conducted for benchmarking the performance of a semi-MRC system \cite{DAF-Liu,DAF-DDF-QZ,DAF-General}. It was shown that the performance of D-AF using semi-MRC is close to the performance of an instantaneous MRC and about 3-4 dB worse than its coherent version \cite{DAF-Liu,DAF-DDF-QZ,DAF-General}.

While obtaining the second-order statistics of  all channels at the destination for combining the received signals could be an issue, the need for a simpler combiner, without sacrificing much of the performance, that can also be analysed exactly, is the motivation of this paper.

In particular, this paper studies D-AF relaying over slow Rayleigh-fading channels using post-detection selection combining (SC) which can be seen as a counterpart to MDPSK for point-to-point communications with reception diversity \cite{SC-DPSK}. At the destination, the decision variable is computed for each link and the one with the maximum magnitude is chosen for non-coherent detection. Hence, different from the semi-MRC, the selection combiner does not need the second-order statistic of any of the channels, which simplifies the destination's detection task. The probability density function (pdf) and commutative density function (cdf) of the instantaneous signal to noise ratio (SNR) in each link and the combiner's output are derived and used to obtain the exact average bit-error-rate (BER) and the outage probability of the system. The analysis is verified with simulation. Comparison of SC and semi-MRC systems shows that the performance of SC is very close to that of the semi-MRC, of course with a lower complexity.

The outline of the paper is as follows. Section \ref{sec:system} describes the system model. In Section III the non-coherent detection of D-AF relaying using SC technique is developed. The performance of the system is considered in Section \ref{sec:symbol_error_probability}. Simulation results are given in Section \ref{sec:sim}. Section \ref{sec:con} concludes the paper.

\emph{Notation}: $(\cdot)^*$, $|\cdot|$ denote conjugate and absolute values of a complex number, respectively. $\mathcal{CN}(0,\sigma^2)$ stand for complex Gaussian distribution with mean zero and variance $\sigma^2$.

\section{System Model}
\label{sec:system}
The wireless relay model under consideration
has one source, one relay and one destination. The source communicates with the destination both directly and via the relay. Each node has a single antenna, and the communication between nodes is half duplex (i.e., each node is able to only send or receive in any given time). The channel coefficients at time $k$, from the source to the destination (SD), from the source to the relay (SR) and from the relay to the destination (RD) are shown with $h_{\mathrm{sd}}[k]$, $h_{\mathrm{sr}}[k]$ and $h_{\mathrm{rd}}[k]$, respectively. All channels are $\mathcal{CN}(0,1)$ (i.e., Rayleigh flat-fading) and follow Jakes' correlation model \cite{microwave-jake}. Also, the channels are spatially uncorrelated and are approximately constant for two consecutive channel uses.

Let $\mathcal{V}=\{{\mathrm{e}}^{j2\pi m/M},\; m=0,\cdots, M-1\}$ be the set of $M$-PSK symbols. A group of $\log_2M$ information bits at time $k$ are transformed to an $M$-PSK symbol $v[k]\in \mathcal{V}$. Before transmission, the symbols are encoded differentially as
\begin{equation}
\label{eq:s-source}
s[k]=v[k] s[k-1],\quad s[0]=1.
\end{equation}
The transmission process is divided into two phases. Block-by-block transmission protocol is utilized to transmit a frame of symbols in each phase as symbol-by-symbol transmission causes frequent switching between reception and transmission, which is not practical.

In phase I,  symbol $s[k]$ is transmitted by the source to the relay and the destination. Let $P_0$ be the average source power per symbol. The received signal at the destination and the relay are
\begin{equation}
\label{eq:source_destination_rx}
y_{\mathrm{sd}}[k]=\sqrt{P_0}h_{\mathrm{sd}}[k]s[k]+w_{\mathrm{sd}}[k]
\end{equation}
\begin{equation}
\label{eq:relay_rx}
y_{\mathrm{sr}}[k]=\sqrt{P_0}h_{\mathrm{sr}}[k]s[k]+w_{\mathrm{sr}}[k]
\end{equation}
where $w_{\mathrm{sd}}[k],w_{\mathrm{sr}}[k]\sim \mathcal{CN}(0,1)$ are noise components at the destination and the relay, respectively.

The received signal at the relay is then multiplied by an amplification factor, and re-transmitted to the destination. The common amplification factor, based on the variance of SR channel, is commonly used in the literature as $A=\sqrt{P_1/(P_0+1)}$, where $P_1$ is the average power per symbol at the relay. However, $A$ can be any arbitrarily fixed value. The corresponding received signal at the destination is
\begin{equation}
\label{eq:dest-rx1}
y_{\mathrm{rd}}[k]=A \; h_{\mathrm{rd}}[k]y_{\mathrm{sr}}[k]+w_{\mathrm{rd}}[k],
\end{equation}
where $w_{\mathrm{rd}}[k]\sim \mathcal{CN}(0,1)$ is the noise at the destination. Substituting (\ref{eq:relay_rx}) into (\ref{eq:dest-rx1}) yields
\begin{equation}
\label{eq:destination-rx}
y_{\mathrm{rd}}[k]= A\; \sqrt{P_0}h[k]s[k]+w[k],
\end{equation}
where $h[k]=h_{\mathrm{sr}}[k]h_{\mathrm{rd}}[k]$ is the equivalent double-Rayleigh channel with zero mean and variance one \cite{SPAF-P} and
$
w[k]=A\; h_{\mathrm{rd}}[k]w_{\mathrm{sr}}[k]+w_{\mathrm{rd}}[k]
$
is the equivalent noise. It should be noted that for a given $h_{\mathrm{rd}}[k]$, $w[k]$ is complex Gaussian random variable with zero mean and variance $A^2 \; |h_{\mathrm{rd}}[k]|^2+1$.

The following section presents the selection combining of the received signals at the destination and its differential detection.


\section{Selection Combining and Differential Detection}
\label{sec:ch-model}
By substituting \eqref{eq:s-source} into (\ref{eq:source_destination_rx}) and (\ref{eq:destination-rx}), and using the slow-fading assumption, $h_{\sd}[k]\approx h_{\sd}[k-1]$ and $h[k]\approx h[k-1]$, one has
\begin{equation}
\label{eq:y_sd_k}
y_{\mathrm{sd}}[k]= v[k] y_{\mathrm{sd}}[k-1]+n_{\mathrm{sd}}[k]
\end{equation}
\begin{equation}
\label{eq:n_sd}
n_{\mathrm{sd}}[k]=w_{\mathrm{sd}}[k]- v[k] w_{\mathrm{sd}}[k-1]
\end{equation}
\begin{equation}
\label{eq:y_rd}
y_{\mathrm{rd}}[k]=v[k] y_{\mathrm{rd}}[k-1]+n_{\mathrm{rd}}[k],
\end{equation}
\begin{equation}
\label{eq:n_rd}
n_{\mathrm{rd}}[k]=w[k]-  v[k] w[k-1].
\end{equation}

Note that, the equivalent noise components $n_{\mathrm{sd}}[k]$ and $n_{\mathrm{rd}}[k]$ (for a given $h_{\mathrm{rd}}[k]$) are  combinations of complex Gaussian random variables, and hence they are also complex Gaussian with variances equal 2 and $2(1+A^2|h_{\rd}[k]|^2)$, respectively.

To achieve the cooperative diversity, the received signals from the two phases should be combined using some combining technique \cite{Linear-Diversity}. For the semi-MRC, the variance of $n_{\sd}$ and the expected value of the variance of $n_{\rd}$ were utilized to combine the signals as \cite{DAF-Liu,DAF-DDF-QZ,DAF-General}
\begin{equation}
\label{eq:y_combined}
\zeta=\frac{1}{2} y_{\mathrm{sd}}^*[k-1]y_{\mathrm{sd}}[k]+\frac{1}{2(1+A^2)} y_{\mathrm{rd}}^*[k-1]y_{\mathrm{rd}}[k]
\end{equation}

\begin{figure}[t]
\psfrag {y1} [] [] [1.0] {$y_{\sd}[k]$}
\psfrag {y2} [] [] [1.0] {$y_{\rd}[k]$}
\psfrag {Delay} [] [] [1.0] {Delay}
\psfrag {Decision} [] [] [1.0] {Selection}
\psfrag {y1k} [l] [] [1.0] {$y_{\sd}^*[k-1]$}
\psfrag {y2k} [l] [] [1.0] {$y_{\rd}^*[k-1]$}
\psfrag {zeta1} [l] [] [1.0] {$\zeta_{\sd}$}
\psfrag {zeta2} [l] [] [1.0] {$\zeta_{\rd}$}
\psfrag {zeta} [l] [] [1.0] {$\zeta$}
\psfrag {*} [] [] [1.0] {*}
\centerline{\epsfig{figure={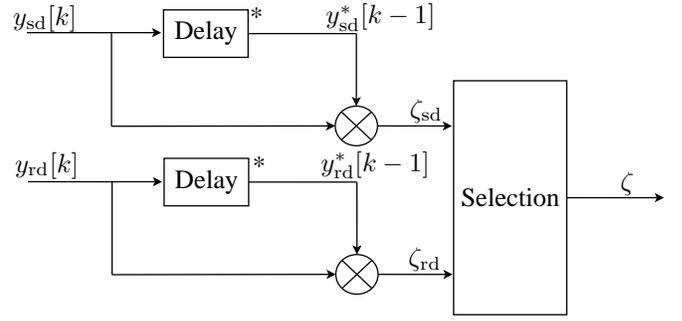},width=8.5cm}}
\caption{Block diagram of the post-detection selection combiner at the destination.}
\label{fig:sc-block}
\end{figure}

However, instead of the semi-MRC which needs the second-order statistics of all channels, we propose to combine the received signals using a selection combiner as illustrated in Figure~\ref{fig:sc-block}. As it is seen, the decision statistics for the direct link, $\zeta_{\sd}=y_{\sd}^*[k-1]y_{\sd}[k]$, and the cascaded link, $\zeta_{\rd}=y_{\rd}^*[k-1]y_{\rd}[k]$, are computed and compared to choose the link with a higher magnitude. The output of the combiner is therefore
\begin{equation}
\label{eq:zeta-sc}
\zeta =
\begin{cases}
\zeta_{\sd} & \mbox{if} \;\; |\zeta_{\sd}|>|\zeta_{\rd}|\\
\zeta_{\rd} & \mbox{if} \;\; |\zeta_{\rd}|>|\zeta_{\sd}|
\end{cases}
\end{equation}
Obviously, no channel information is needed at the destination.

Finally, the well known minimum Euclidean distance (ED) detection is applied to detect the transmitted signal as \cite{Dig-comm-porakis}
\begin{equation}
\label{eq:ml-detection}
\hat{v}[k]= \arg \min \limits_{x\in \mathcal{V}} |\zeta-x|^2,
\end{equation}
where the minimization is taken over all symbols $x$ of the constellation $\mathcal{V}$.

In the next section, the performance of the above selection combining detector is analysed.

\section{Error Performance Analysis}
\label{sec:symbol_error_probability}
In order to evaluate the performance of the system, the distribution of the instantaneous SNR at the output of the selection combiner is derived and used in the unified approach \cite{unified-app} to obtain the BER. To simplify the notation, the time index of the channels is omitted in this section.

The instantaneous received SNRs of two links are given as \cite{DAF-Liu,DAF-DDF-QZ,DAF-General}
\begin{gather}
\label{eq:gama_sd}
\gamma_{\sd}=P_0 |h_{\sd}|^2\\
\label{eq:gama_rd}
\gamma_{\rd}= c |h_{\sr}|^2
\end{gather}
where $c=A^2 P_0 |h_{\rd}|^2/(1+A^2 |h_{\rd}|^2)$. Since, $|h_{\sd}|^2$ has an exponential distribution, $\gamma_{\sd}$ is also exponentially distributed with the following pdf and cdf:
$
\label{eq:pdf-gama-sd}
f_{\gamma_{\sd}}(\gamma)=(1/P_0) e^{-\frac{\gamma}{P_0}}
$,
$
\label{eq:cdf-gamma-sd}
F_{\gamma_{\sd}}(\gamma)=1-e^{-\frac{\gamma}{P_0}}.
$

Since, the quantity $c$ conditioned on $h_{\rd}$ is a constant, the conditional pdf and cdf of $\gamma_{\rd}$ are given as
$
\label{eq:pdf-gama-rd}
f_{\gamma_{\rd}|h_{\rd}}(\gamma)=(1/c) e^{-\frac{\gamma}{c}}
$,
$
\label{eq:cdf-gama-rd}
F_{\gamma_{\rd}|h_{\rd}}(\gamma)=1-e^{-\frac{\gamma}{c}}.
$

The instantaneous SNR at the output of the combiner is defined as
$
\label{eq:gama-max}
\gamma_{\max}=\max (\gamma_{\sd},\gamma_{\rd})
$
Thus, its cdf, conditioned on $h_{\rd}$, can be expressed as
\begin{equation}
\label{eq:cdf-gama-max}
\begin{split}
F_{\gamma_{\max}|h_{\rd}}(\gamma)&= \mbox{Pr}(\max(\gamma_{\sd},\gamma_{\rd}) \leq \gamma |h_{\rd} )\\
&=\mbox{Pr}(\gamma_{\sd}\leq \gamma,\gamma_{\rd} \leq \gamma | h_{\rd} )\\
&= F_{\gamma_{\sd}}(\gamma) F_{\gamma_{\rd}|h_{\rd}}(\gamma)\\
&=\left( 1-e^{-\frac{\gamma}{P_0}}\right) \left(1-e^{-\frac{\gamma}{c}}\right)
\end{split}
\end{equation}
By taking the derivative of \eqref{eq:cdf-gama-max}, the conditional pdf of $\gamma_{\max}$ is
\begin{equation}
\label{eq:pdf-gama-max}
f_{\gamma_{\max}|h_{\rd}}(\gamma)=\frac{1}{P_0} e^{-\frac{\gamma}{P_0}}+ \frac{1}{c} e^{-\frac{\gamma}{c}}-\frac{1}{c'} e^{-\frac{\gamma}{c'}}
\end{equation}
where $c'=c P_0/(c+P_0)$ conditioned on $h_{\rd}$ is a constant.

Using the unified approach \cite[eq.25]{unified-app}, it follows that the conditional BER can be written as
\begin{equation}
\label{eq:Pb-gama-hrd}
P_b(E|\gamma_{\max},h_{\rd})=\frac{1}{4\pi} \int \limits_{-\pi}^{\pi} g(\theta) e^{-\alpha(\theta)\gamma_{\max}} \dd \theta
\end{equation}
where $g(\theta)=(1-\beta^2)/(1+2\beta\sin(\theta)+\beta^2)$, $\alpha(\theta)=(b^2/(2\log_2 M))(1+\beta^2+2\beta\sin(\theta))$, and $\beta=a/b$. The values of $a$ and $b$ depend on the modulation size \cite{unified-app}.

Next, the average over the distribution of $\gamma_{\max}$ is taken to give
\begin{equation}
\label{eq:Pb-hrd}
P_b(E|h_{\rd})=\frac{1}{4\pi} \int \limits_{-\pi}^{\pi} \int \limits_{0}^{\infty} g(\theta) e^{-\alpha(\theta)\gamma} f_{\gamma_{\max}|h_{\rd}}(\gamma) \dd \gamma \dd \theta
\end{equation}
By substituting \eqref{eq:pdf-gama-max} into \eqref{eq:Pb-hrd}, one obtains
\begin{equation}
\label{eq:Pb-hrd-sub}
P_b(E|h_{\rd})=\frac{1}{4\pi} \int \limits_{-\pi}^{\pi} g(\theta) [I_1(\theta)+I_2(\theta)-I_3(\theta)] \; \dd \theta
\end{equation}
where
\begin{equation}
\label{eq:I1-theta}
I_1(\theta)=\int \limits_{0}^{\infty} e^{-\alpha(\theta)\gamma} \frac{1}{P_0} e^{-\frac{\gamma}{P_0}} \dd \gamma
= \frac{1}{P_0 \alpha(\theta)+1}
\end{equation}

\begin{equation}
\label{eq:I2-theta}
I_2(\theta)=\int \limits_{0}^{\infty} e^{-\alpha(\theta)\gamma} \frac{1}{c} e^{-\frac{\gamma}{c}} \dd \gamma
= \frac{1}{c \alpha(\theta)+1}
\end{equation}
\begin{equation}
\label{eq:I3-theta}
I_3(\theta)=\int \limits_{0}^{\infty} e^{-\alpha(\theta)\gamma} \frac{1}{c'} e^{-\frac{\gamma}{c'}} \dd \gamma
= \frac{1}{c' \alpha(\theta)+1}
\end{equation}

Finally, substituting $c$ and $c'$ and taking the average over the distribution of $|h_{\mathrm{rd}}|^2$, $f_{\lambda}(\lambda)=e^{-\lambda},\hspace{.1 in} \lambda>0$, the unconditioned BER is given as
\begin{equation}
\label{eq:BER-integral}
P_b(E)=\frac{1}{4\pi} \int \limits_{-\pi}^{\pi} g(\theta)[J_1(\theta+J_2(\theta)-J_3(\theta)] \; \dd \theta
\end{equation}
where
\begin{equation}
\label{eq:J1-theta}
J_1(\theta)=\int \limits_{0}^{\infty} I_1(\theta) e^{-\lambda} \dd \lambda\\
= \frac{1}{P_0 \alpha(\theta)+1}
\end{equation}
\begin{equation}
\label{eq:J2-theta}
\begin{split}
J_2(\theta)&=\int \limits_{0}^{\infty} I_2(\theta) e^{-\lambda} \dd \lambda\\
&= b_3(\theta) [1+(b_1-b_2(\theta))e^{b_2(\theta)}E_1(b_2(\theta))]
\end{split}
\end{equation}
with $b_1=1/A^2$, $b_2(\theta)=1/(A^2(1+P_0\alpha(\theta))$ and $b_3(\theta)=1/(P_0\alpha(\theta)+1)$.
\begin{equation}
\label{eq:J3-theta}
\begin{split}
J_3(\theta)&=\int \limits_{0}^{\infty} I_3(\theta) e^{-\lambda} \dd \lambda\\
&= d_3(\theta) [1+(d_1-d_2(\theta))e^{d_2(\theta)}E_1(d_2(\theta))]
\end{split}
\end{equation}
with $d_1=1/(2A^2)$, $d_2(\theta)=1/(A^2(2+P_0\alpha(\theta))$ and $d_3(\theta)=2/(P_0\alpha(\theta)+2)$.
Also, $E_1(x)=\int \limits_x^{\infty} (e^{-t}/t)\dd t$ is the exponential integral function.
The integral in \eqref{eq:BER-integral} can be computed numerically to find the exact BER.

To get more insights about the achieved diversity, approximating $J_2(\theta)$ and $J_3(\theta)$ with $b_3(\theta)$ and $d_3(\theta)$ as
\begin{equation}
\label{eq:upper-bounds}
\begin{split}
J_2(\theta) & \gtrsim \frac{1}{\alpha(\theta)P_0+1},\\
J_3(\theta) & \gtrsim \frac{2}{\alpha(\theta)P_0+2}
\end{split}
\end{equation}
Using the above values in \eqref{eq:BER-integral}, it can be seen that
\begin{equation}
\label{eq:approx-ub}
P_b(E)\propto \frac{2}{(1+\alpha(\theta)P_0)(2+\alpha(\theta)P_0)} \propto \frac{1}{P_0^2}
\end{equation}
which shows that the diversity order of two can be achieved in high SNR region.

Before closing this section, it is pointed out that the outage probability can be straightforwardly obtained from \eqref{eq:pdf-gama-max}. Specifically, the probability that the instantaneous SNR at the output of the SC combiner drops below a SNR threshold $\gth$ is
\begin{multline}
\label{eq:Poutage}
P_{\mathrm{out}}=\mathrm{Pr}(\gamma_{\max}\leq \gth)
= \int \limits_{0}^{\infty} F_{\gamma_{\max}|h_{\rd}} (\gth) e^{-\lambda} \dd \lambda \\=
\left(1-e^{-\frac{\gth}{P_0}}\right)\left[1- e^{-\frac{\gth}{P_0}} \sqrt{\frac{4\gth}{A^2P_0}}K_1\left(\sqrt{\frac{4\gth}{A^2P_0}}\right)\right]
\end{multline}
where $K_1(\cdot)$ is the first-order modified Bessel function of the second kind.

\section{Simulation Results}
\label{sec:sim}
To verify the BER performance analysis, computer simulation was carried out.\footnote{Due to space limitation, simulation results that verify the outage probability analysis are not included.} In the simulation, the channels $h_{\mathrm{sd}}[k]$, $h_{\mathrm{sr}}[k]$ and $h_{\mathrm{rd}}[k]$ are generated individually according to the simulation method of \cite{ch-sim}. The normalized Doppler frequency of all channels is set to $0.001$, so that the channels are slow-fading. Binary data is differentially encoded for $M=2,\;4$ constellations. At the destination, the received signals are combined using the SC technique and the decision variable is used to recover the transmitted signal using the minimum Euclidean-distance detection. The simulation is run for various values of the total power in the network, whereas the amplification factor at the relay is fixed to $A=\sqrt{P_1/(P_0+1)}$ to normalized the average relay power to $P_1$. 

\begin{figure}[t!]
\psfrag {BER} [] [] [1] {BER}
\psfrag {c} [t] [b] [1] {$q=\frac{P_0}{P}$}
\centerline{\epsfig{figure={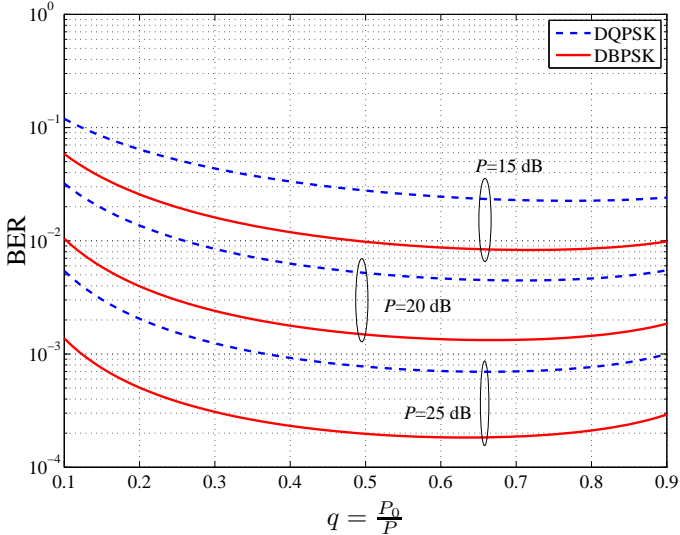},width=9cm}}
\caption{BER as a function of power allocation factor $q$ for $P=15,\;20,\;25$ dB.}
\label{fig:pw_all}
\end{figure}

First, to find the optimum power allocation between the source and the relay, the expression of BER is examined for different values of power allocation factor $q=P_0/P$, where $P=P_0+P_1$ is the total power in the system. The BER curves are plotted versus $q$ in Figure~\ref{fig:pw_all} for P=15,\;20,\;25 dB and when DBPSK and DQPSK are employed. Note that, for computing the theoretical BER in \eqref{eq:BER-integral}, $\left\lbrace a=0,\; b=\sqrt{2}\right\rbrace$ and $\left\lbrace a=\sqrt{2-\sqrt{2}},\; b=\sqrt{2+\sqrt{2}}\right\rbrace$ are obtained for DBPSK and DQPSK, respectively \cite{unified-app}.
The figure shows that more power should be allocated to the source than the relay and the BER is minimized at $q\approx 0.7$. This observation is similar to what reported in \cite{DAF-Liu} for the semi-MRC technique. Based on Figure~\ref{fig:pw_all} the power allocation factor $q=0.7$ is used in all the simulations.

\begin{figure}[b]
\psfrag {P(dB)} [][] [1]{$P$ (dB)}
\psfrag {BER} [] [] [1] {BER}
\centerline{\epsfig{figure={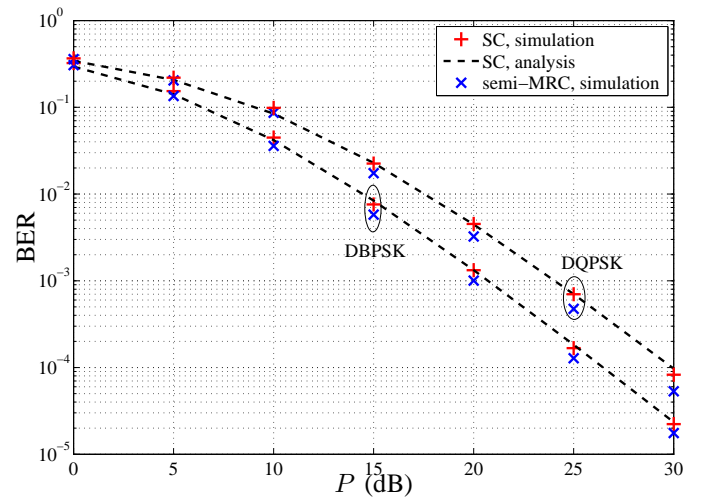},width=9cm}}
\caption{Theoretical and simulation BER of the D-AF system with semi-MRC and SC methods using DBPSK (lower) and DQPSK (upper).}
\label{fig:sc_m2m4}
\end{figure}

Figure~\ref{fig:sc_m2m4} plots the BER curves versus the total power $P$ that are obtained with the SC technique (both theoretical and simulation results) and the semi-MRC technique, and for both DBPSK (lower plots) and DQPSK (upper plots). As can be seen, the simulation results of SC technique are very close to the theoretical values. Moreover, the diversity order of two is achieved for both SC and semi-MRC methods and their results are also very close to each other. The small difference between the two methods can be accepted in many practical applications which seek a trade-off between simplicity and performance.

\balance

\section{Conclusion}
\label{sec:con}
A selection combining of the received signals at the destination of a D-AF relay network was studied. Thanks to the differential encoding and selection combiner, no channel state information is needed at the destination for detection of the transmitted symbols. The distribution of the instantaneous SNR at the output of the combiner was derived and the exact bit error rate and the outage probability of the system have been obtained. It was shown that the desired diversity order of two can be achieved by the SC system. Simulation results verified the analysis and show that the selection combiner performs very close to the more-complicated semi-MRC technique (which needs the second-order statistics of all channels).

\bibliographystyle{IEEEbib}
\bibliography{ref/references}

\end{document}